\newcommand{\be}{\begin{equation}}
\newcommand{\ee}{\end{equation}}
\newcommand{\bea}{\begin{eqnarray}}
\newcommand{\eea}{\end{eqnarray}}

\documentstyle[12pt]{article}
\advance\textheight by \topskip
\begin{document}

\begin{flushright}
SU-ITP-97-36\\

hep-th/9709069\\
Sptember 9, 1997
\end{flushright}
\vspace{.5cm}

\begin{center}
\baselineskip=16pt

{\Large\bf  Worldvolume Supersymmetry} \\

\

\vskip 1 cm

{\bf  Renata Kallosh}

\vskip 1cm

{\em Physics Department, Stanford
University, Stanford, CA 94305-4060, USA\\
kallosh@physics.stanford.edu
}

\end{center}

\vskip 1 cm
\centerline{\bf ABSTRACT}
\vspace{-0.3cm}
\begin{quote}

A general set of rules is given  how to convert a local $\kappa$-symmetry of a
brane action and space-time supersymmetry into the global supersymmetry of the
worldvolume. A Killing spinor adapted gauge for quantization of
$\kappa$-symmetry is defined for this purpose.
As an application of these rules we perform the gauge-fixing of the M-5-brane
to get the   theory of the (0, 2)  tensor supermultiplet in d=6.
\end{quote}
\normalsize

\baselineskip=16pt

\newpage
\section{Introduction}
The set of known $\kappa$-symmetric actions include p-branes and more recent
D-p-branes\cite{Ce1,Ag1,Be1}  and M-5-brane  \cite{5b,ag} .
The worldvolume fields of $\kappa$-symmetric p-brane actions comprise the maps
$Z^M (\xi^i)$ from the worldvolume coordinates $\xi^i = 0, \dots , p$ to the
superspace $Z^M= \Bigl (X^{ m}  , \theta^ \mu \Bigr )$.
D-p-branes depend in addition on world-volume  1-form gauge potential and
M-5-brane
depends on a 2-form gauge potential with the self-dual field strength.  All
these actions in a flat background have  a global space-time supersymmetry
as well as a local $\kappa$-symmetry which is in general infinite reducible and
difficult to deal with. The quantization was developed mostly in the light-cone
gauge for the Green-Schwarz string and for the Bergshoeff-Sezgin-Townsend
membrane. Recently it became possible in case of D-p-branes to define an
irreducible  $\kappa$-symmetry  and exhibit the worldvolume supersymmetry upon
gauge-fixing an irreducible   $\kappa$-symmetry \cite{kallosh} . The
gauge-fixing of D-p-branes was possible in a covariant way with respect to a
10-dimensional Lorentz symmetry \cite{Ag1,bk, kallosh}.

The purpose of this note is to introduce the concept of irreducible
$\kappa$-symmetry and its consequent gauge-fixing for  a general case of
$\kappa$-symmetric actions. We are not trying to achieve here the quantization
covariant in embedding space-time, in general,  although the experience with
covariant quantization of D-p-branes is helpful.
For the M-branes as different from D-p-branes  this type of covariant
quantization is certainly not possible as the 32 component Majorana spinor
forms the smallest representation of the 11-dimensional Lorentz group.

Our main purpose here is  to find the best way to exhibit the worldvolume
global supersymmetry of the brane. In particular we would like to get the
supersymmetric action and global supersymmetry transformation rules for the (0,
2)  tensor multiplet in  d=6 theory. Not very much is known about the tensor
multiplets in d=6.
In the case of (2,0) supersymmetry, the equations of motion describing the
coupling of $n$ tensor multiplets to supergravity have been constructed
\cite{Romans1}. In the case of (1,0) supersymmetry the coupling of the tensor
multiplet to Yang-Mills multiplet in absence of supergravity, is known
\cite{BSS}.
The interest to the 6-dimensional supersymmetric theory of a tensor multiplet,
besides that it is a gauge-fixed M-5-brane theory,  is also motivated by the
expectation that in Matrix theory \cite{BFSS} the theory of
$n$ interacting (0, 2)  tensor supermultiplets may play an important role
\cite{S}.

We will use both the action of the M-5-brane \cite{5b} as well as the geometric
approach to M-5-brane developed in \cite{HSW} based on superembedding the
target superspace into the worldvolume superspace of the brane.

The relation between space-time supersymmetry, $\kappa$-symmetry
and unbroken worldvolume supersymmetry was established in \cite{bk}. The
unbroken worldvolume supersymmetry  defining the BPS states on the brane was
found to be given by a universal formula
\begin{equation}
\delta \theta_{\rm unbr} = (1-\Gamma) \epsilon =0
\end{equation}
where $(1+\Gamma)$ is the generator of $\kappa$-symmetry $\delta_{\kappa}
\theta = (1+\Gamma) \kappa $.  Also  the algebra of Noether supercharges of the
M-5-brane classical action was studied in \cite{st}.

In this note we will find the worldvolume supersymmetry which appears in the
gauge-fixed action of the M-5-brane.

The strategy is to adapt here  the rules of duality symmetric quantization in
\cite{duality}
where it was  suggested to {\it gauge-fix the infinite reducibility of the
$\kappa$-symmetry
using the Killing spinors admitted by a consistent background of a given
extended object}. This will give us the suitable way to get the worldvolume
supersymmetry. However we will not restrict ourself with these class of gauges
only. In fact we will work out  the general class of gauges which will serve
the purpose of quantization of $\kappa$-symmetry. It is expected that the
physical matrix elements of the theory are independent  on the choice of the
gauge. However the global symmetries of the theories may take
different form in different gauges. This freedom will be used for the  simplest
possible description of the worldvolume supersymmetric theories.

\section{Irreducible $\kappa$-symmetry}
The class of actions we consider  have 32-dimensional
global space-time supersymmetry and the local $\kappa$-supersymmetry are :
\begin{equation}
\delta_\epsilon \theta = \epsilon, \qquad \delta_\epsilon X^m = \bar\epsilon
\Gamma^m
\theta \ , \label{susytrans}
\end{equation}
\begin{equation}
\delta \theta = (1+\Gamma)  \kappa \ , \qquad
 \delta X^m = \bar\theta \Gamma^m \delta\theta \label{Xtrans}  \ , \dots \ .
\end{equation}
Here dots mean the transformations of Born-Infeld or a tensor field. $ \Gamma$
is a function of the fields of the brane and depends on $\xi^i$ therefore. The
matrix  $\Gamma(\xi)$  squares to 1 and has a vanishing trace:
\begin{equation}
{\rm tr}\ \Gamma=0\, ,
\hspace{1cm}
\Gamma^2=1  \ ,\qquad (1+ \Gamma) (1- \Gamma)=0\, ,
\end{equation}
i.e.  $1+ \Gamma$ is a projector which makes a 32-dimensional parameter of
$\kappa$-supersymmetry effectively only 16-dimensional.  Let us pick up some
constant, $\xi^i$-independent projectors
\begin{equation}
{\cal P_{\pm} } = {1\over 2} \Bigl( 1\pm \gamma \Bigr) \end{equation}
which can divide any 32-dimensional spinor  into 2 parts. Here again we assume
that
\begin{equation}
{\rm tr}\ \gamma=0\, ,
\hspace{1cm}
\gamma^2=1\, , \qquad (1+ \gamma) (1- \gamma)=0\, .
\end{equation}
This $\xi^i$-independent projector will be used to fix the gauge.
We will call the gauge adapted to the Killing spinor  when
\begin{equation}
\gamma=  \Gamma|_{\rm cl} \, ,
\end{equation}
i.e. the constant projector $1+ \gamma$ is the $\kappa$-symmetry projector  $1+
\Gamma$ taken at the values of fields which form a classical solution
describing the relevant bosonic brane. The Killing spinor of a space-time
geometry naturally can not depend on the coordinates of the worldvolume.
We will see it in an  example of a M-5-brane later. In general we do not
require any relations between  possible projectors for gauge-fixing and
$\kappa$-symmetry generators. We will find out later some constraints which
are required to make a projector defining the gauge-fixing possible.

The steps to gauge-fix $\kappa$-symmetry and get the worldvolume global
supersymmetry in the most general case are:

\begin{enumerate}

\item Find the basis in which $\gamma$ is diagonal so that

\begin{equation}
{\cal P_+ }  = \pmatrix{1&
0 \cr
\cr              0 &0}\, , \qquad {\cal P_- }  = \pmatrix{0&
0 \cr
\cr              0 &1}\, .
 \end{equation}
and split all spinors accordingly:
\begin{equation}
\theta = \left (\matrix{
\theta ^\alpha \cr
\theta ^{\alpha'} \cr
}\right ) \ ,  \qquad \kappa = \left (\matrix{
\kappa ^\alpha \cr
\kappa ^{\alpha'} \cr
}\right ) \ , \qquad \epsilon = \left (\matrix{
\epsilon ^\alpha \cr
\epsilon ^{\alpha'} \cr
}\right )
\end{equation}

\item In this basis define the block structure of $1\pm \Gamma$ as follows
\begin{equation}
1 + \Gamma = \pmatrix{1+C&
(1- C^2) A^{-1}\cr
\cr                A &1- C}\, , \qquad 1 - \Gamma = \pmatrix{1-C&
-(1- C^2) A^{-1}\cr
\cr                -A &1+ C}\
\end{equation}
where the $16\times 16$ dimensional matrices $C$ and $A$ commute.
\begin{equation}
AC-CA=0
\end{equation}
The matrices  $1 \pm\Gamma$ have  vanishing determinants and rank 16, which
means that the 16-dimensional matrices $1\pm C$ and $A$ are invertible.

\item If some  choice of a projector ${\cal P_\pm }$ leads to non-invertible A
or
$1\pm C$, this projector   can not be used for quantization. Examples include
d=10 covariant gauges for the Green-Schwarz string.

\item Introduce the irreducible  16-dimensional $\kappa$-symmetry by the
constraint
\begin{equation}
{\cal P_- } \kappa = {1\over 2} (1- \gamma) \kappa =0
\end{equation}
In our basis this means that 16 components of $\kappa$-symmetry vanish
\begin{equation}
 \kappa = \left (\matrix{
\kappa ^\alpha \cr
0 \cr
}\right ) \qquad \kappa ^{\alpha'} =0
\end{equation}
Irreducible $\kappa$-symmetry takes the form
\begin{eqnarray}
\delta_{\kappa} \theta ^\alpha  &=& (1+C)^\alpha{}_\beta \kappa^\beta
\nonumber\\
\nonumber\\
\delta_{\kappa} \theta ^{\alpha'} &=& A ^{\alpha'}{}_\beta \kappa^\beta
\end{eqnarray}

\item Consider a combination of 16-dimensional irreducible $\kappa$-symmetry
and 32-dimensional space-time supersymmetry
\begin{eqnarray}
\delta_{\kappa ,  \epsilon} \theta ^\alpha  &=& (1+C)^\alpha{}_\beta
\kappa^\beta + \epsilon ^\alpha \nonumber\\
\nonumber\\
\delta_{\kappa ,  \epsilon'}  \theta ^{\alpha'} &=& A ^{\alpha'}{}_\beta
\kappa^\beta + \epsilon^{\alpha'}
\end{eqnarray}

\item Fix the 16-dimensional irreducible $\kappa$-symmetry by imposing 16 gauge
conditions
\begin{equation}
{\cal P_+ } \theta = {1\over 2} (1+ \gamma) \theta =0  \qquad  \Longrightarrow
\qquad \theta = \left (\matrix{
0 \cr
\theta^{\alpha'} \cr
}\right ) \qquad \theta ^{\alpha} =0
\end{equation}
 \item  Find the relation between $\kappa$ and $\epsilon$ which will keep the
gauge $\theta ^{\alpha} =0$.
\begin{equation}
 \kappa^\beta =- [(1+C)^{-1}] ^\beta{}_ \alpha \epsilon ^\alpha
\end{equation}

\item Finally get the 32-dimensional supersymmetry
transformation\footnote{If the reparametrization symmetry is fixed by choosing
a static gauge, the space-time spinors (former scalars on the worldvolume)
like  $\theta ^{\alpha'}$ and
$\epsilon ^\alpha , \epsilon^{\alpha'}
$ become worldvolume spinors.}
of  16  $ \theta ^{\alpha'}$ living on the brane
\begin{equation}
\delta_{\epsilon , \epsilon'}  \theta ^{\alpha'} = - A ^{\alpha'}{}_\beta
[(1+C)^{-1}] ^\beta{}_ \alpha \epsilon ^\alpha + \epsilon^{\alpha'}
\end{equation}

This is the {\it general answer}. Given a $\kappa$-symmetry of the action is
known and the right choice of the constant projector is made, which supplies us
with $16\times 16$ matrices $C$ and $A$, we have the answer for the worldvolume
supersymmetry.

\end{enumerate}

For example for D-p-branes we have for $p$ even $\gamma= \Gamma_{11}$,  $C=0$
and an invertible $A$ can be find in \cite{Ag1,bk,kallosh} together with the
total procedure, described here  in steps 1-8. This is an example when both
$1+C$ and $A$ are invertible for a given choice of $\gamma$.
With the same choice of $\gamma$  type IIA GS string will have a non-invertible
$A=0$ and this gauge is not acceptable as one can verify.

\section{ The theory of the (0, 2)  tensor supermultiplet in d=6}
We will give here a brief description of gauge-fixing procedure of the
M-5-brane theory which provides the supersymmetric action for (0, 2)  tensor
supermultiplet in d=6. The Padova-Kharkov  manifestly d=6 general coordinate
invariant M-5-brane action is  \cite{5b}
\be
S_{M-5} \Bigl (X^m (\xi), \theta^\mu(\xi), A_{jk} (\xi) , a (\xi) \Bigr )=
\int\! d^6\xi\, (L_0 + L_{WZ})\, ,
\ee
where
\bea
\label{23}
L_0 &=& -\sqrt{-\det(g_{ij}+{\tilde H}_{ij})}
+ {\sqrt{-g} \over 4(\partial a \cdot \partial a)}
({\partial}_i a) (H^*)^{ijk} H_{jkl}({\partial}^l a) \label{23a}\\
L_{WZ} &=& {1\over{6!}}\varepsilon ^{i_1\dots i_6}
\big[C^{(6)}_{i_1\dots i_6}+
10 H_{i_1i_2i_3} C^{(3)}_{i_4i_5i_6}\big] \, .
\eea
Here
$g=\det(g_{ij})$, and \be \label{21} (H^*)^{ijk} = {1\over 3!\sqrt{-g}}
\varepsilon^{ijki'j'k'}H_{i'j'k'} \, ,\qquad {\tilde H}^{ij} = {1\over
\sqrt{-(\partial a \cdot \partial a)}}\, (H^*)^{ijk}\partial_k a\, , \ee
The generalized field strength of the tensor field is
\be
\label{19}
H_{ijk}={\partial}_{[i}A_{jk]}-C^{(3)}_{ijk}\, ,
\ee
where $C^{(3)}_{ijk}$ is the pullback of the superspace 3-form gauge potential
$C^{(3)}$. The induced worldvolume
metric $g_{ij}(\xi )= E_i{}^a E_j{}^b \eta_{ab}$, where $\eta$ is the D=11
Minkowski metric and $E_i{}^a =\partial_i Z^M E_M{}^a$. The worldvolume
six-form $C^{(6)}_{i_1\dots i_6}$ is
induced by the superspace 6-form gauge potential. The auxiliary worldvolume
scalar field $a(\xi)$  serves to achieve the manifestly d=6 general coordinate
invariance of the M-5-brane action.
The $\kappa$-symmetry transformations and supersymmetry of space-time fermions
are
\begin{equation}
\delta_{\kappa, \epsilon } \theta = (1+ \Gamma)\kappa + \epsilon
\end{equation}
where
\begin{equation}
\Gamma=\Gamma_{(0)}  + \Gamma_{(3)}\,  \\
\end{equation}
and
\begin{eqnarray}
\Gamma_{(0)}&=& {\textstyle\frac{1}{6! \sqrt{|g|}}} \epsilon^{i_1\cdots i_6}
\gamma_{i_1}\cdots
\gamma_{i_6}\  \qquad \Gamma_{(3)}={1\over 2\cdot 3!} h_{ijk} \gamma^{ijk}
\nonumber\\
\nonumber\\
i&=&0,\dots ,5 \ ,  \qquad m=0,\dots , 10 \ ,  \qquad \gamma_i = E_i{}^m
\Gamma_m \ .
\end{eqnarray}
Here we are using the form of
$\kappa$-symmetry transformations found originally in
the superembedding approach \cite{HSW} and proved later to be also a symmetry
transformation of the M-5-brane action in \cite{equiv}. The
worldvolume field $h_{ijk}$  of \cite{HSW} turns out to be   a non-linear
function of the fields in the action, whose explicit form can be found in
\cite{equiv}.
Note that due to the self-duality of $h$  and the nilpotency of $
\Gamma_{(3)}$,  $\Gamma$ can also be given in a form
\cite{bk}
\begin{equation}
\Gamma= e^{ \Gamma_{(3)}}\Gamma_{(0)}\, ,
 \qquad  (\Gamma_{(3)})^2=0
\end{equation}

The gauge-fixing of the M-5-brane action is inspired by the superembedding
\cite{HSW} of the space-time superspace with coordinates $X^m, \Theta^\mu$ into
worldvolume superspace with coordinates $\xi^i, \theta^\alpha$. We split $m=(i,
a'), \mu = (\alpha, \alpha')$. The superembedding is $X^i = \xi^i \ ,
\Theta^\alpha = \theta^\alpha$ and $X^{a'} =X^{a'}(\xi, \theta) \ ,
\Theta^{\alpha'}= \Theta^{\alpha'}(\xi, \theta)$. To be as close to this as
possible in the bosonic action of the 5-brane we have to require that in our
action
\begin{equation}
X^i = \xi^i, \; \theta^\alpha=0
\label{gauge}\end{equation}
and the fields of the (0,2) tensor multiplet remaining in the action which
depend on $\xi$  are
\begin{equation}
X^{a'}(\xi ) , \; \theta^{\alpha'}(\xi) , A_{ij}(\xi )  , a(\xi ) \ , \qquad
a'=1,2,3,4,5 \ , \qquad \alpha' = 1,2,\dots ,16.
\end{equation}
Thus we have 5 scalars $X^{a'}(\xi )$, a 16-component spinor $\;
\theta^{\alpha'}(\xi)$ which can considered (see below) as a  a chiral d=6
spinor with a $USp(4)$ symplectic  Majorana-Weyl reality condition
$\theta^{\hat \alpha}_s$, a tensor $A_{ij}(\xi ) $ with the self-dual field
strength and an auxiliary scalar $a(\xi )$.
The 11d  $32\times 32$ $\Gamma^m$ matrices have to be taken in the basis which
correspond to the split of the target superspace into the superspace of the
5-brane and the rest \cite{HSW}.  This reflects the
 $Spin (1,5)\times USp(4)$ symmetry of the six dimensional theory.  An 11d
Majorana spinor decomposes as
\begin{equation}
 \psi = (\psi_{\hat \alpha s} , \; \psi^{\hat \alpha}_s)
\end{equation}
where $s=1,2,3,4$ is an $USp(4)$ index and $\hat \alpha=1,2,3,4$ is a 6d Weyl
spinor index with upper (lower) indices corresponding to anti-chiral (chiral)
spinors respectively. The 6d spinors satisfy a Majorana-Weyl reality condition.
The relevant representation of 11d   $\Gamma^m$ is
\begin{equation}
\Gamma^i _{\hat \alpha s, \hat \beta t} = \eta_{st} (\sigma^i)_{\hat \alpha
\hat \beta}
\end{equation}
where $\eta_{st}$ is the $USp(4)$ antisymmetric invariant metric and $\sigma^i$
the 6d chirally-projected gamma-matrices etc. \cite{HSW}. In terms of
16-component spinors we have $ \psi_{\hat \alpha s}=\psi ^{\alpha' }, \;
\psi^{\hat \alpha}_s= \psi^{\alpha}.
$

Thus  we choose a projector $\gamma$ to be  a chiral projector of the
6-dimensional space times the unit matrix. In the basis above this means that
\begin{equation}
\gamma= \Gamma|_{cl} = {\textstyle\frac{1}{6! }} \epsilon^{i_1\cdots i_6}
\gamma_{i_1}\cdots
\gamma_{i_6}|_{cl}  = \pmatrix{
1 &  0 \cr
0 & \ -1 \cr
}\
\end{equation}
 Here the subscript ${}_{cl} $ means that we take $X^{a'}_{cl}={\rm const}, \;
\theta_{cl}=0, (h_{ijk})_{cl} =0,  (E_{i} {}^j)_{cl}= \delta _i{}^j,  (E_{ i}
{}^{a'})_{cl}=0 $ and the field-independent part of the $\kappa$-symmetry
generator $1+ \Gamma$ provides us with the projector for gauge-fixing
$\kappa$-symmetry. Note that this is exactly the projector which specifies the
M-5-brane Killing spinor in the target space.\footnote{ An interesting
possibility suggested by E. Bergshoeff  of another gauge-fixing the M-5-brane
will use the new classical solution of Howe, Lambert and West \cite{HSW}, with
$(h_{ijk})_{cl} \neq 0$  describing the self-dual string soliton of the
M-5-brane. Our quantization procedure may require modification when the  BPS
classical solutions on the brane are used for projectors.}

The gauge fixed theory is given by the classical action in the gauge
(\ref{gauge}). Since both the reparametrization symmetry  and $\kappa$-symmetry
are fixed in a unitary way, there are no propagating ghosts. We do not
gauge-fix the Maxwell theory of the tensor fields as our main purpose is to get
the full theory of  the (0, 2)  tensor supermultiplet in d=6.

{\it The action for a tensor multiplet is an action of a M-5 brane in a Killing
spinor adapted gauge with vanishing $X^i-\xi^i$ and $  \theta^\alpha (\xi) $}:

\be
S_{(0,2)}  \Bigl (X^{a'}  (\xi), \theta^{\alpha'} (\xi), A_{jk} (\xi) , a (\xi)
\Bigr )=
S_{M-5} \Bigl (X^{a'} (\xi) ,  \theta^{\alpha'} (\xi), A_{jk} (\xi) , a (\xi) ,
X^i-\xi^i=0 , \theta^\alpha (\xi) =0 \Bigr )
\ee

To find   the exact non-linear worldvolume supersymmetry transformations of the
$S_{(0,2)}$  action
we may now proceed using the rules from the previous section. In addition to
gauge fixing the spinor theta we have to gauge fix the infinite reducible
$\kappa$-symmetry. We choose as before
\begin{equation}
\theta^\alpha=0 \qquad \kappa^{\alpha'} =0
\end{equation}
To extract from the generator of  $\kappa$-symmetry $\Gamma$ the matrices $C$
and $A$ which define the worldvolume supersymmetry we have to take into account
that  in the flat 11-dimensional background
\begin{equation}
\Gamma= {\textstyle\frac{1}{6! \sqrt{|g|}}} \epsilon^{i_1\cdots i_6} (
\gamma_{i_1}\cdots
\gamma_{i_6}\   + 40 \gamma_{i_1} \gamma_{i_2}
\gamma_{i_3} h_{i_4 i_5 i_6}) \ , \\
\end{equation}
\begin{eqnarray}
 \gamma_i = (\delta_i{}^j - i \bar \theta \Gamma^j \partial_i \theta )
\Gamma_j+ \partial _i X^{a'} \Gamma_{a'}\ .
\label{gamma}\end{eqnarray}
Here we used the fact that the spinors are chiral and therefore $\bar \theta
\Gamma^{a'} \partial_i \theta $ vanishes. Using eq. (\ref{gamma}) we may
rewrite $\Gamma$ as a sum of products of $\Gamma$'s
\begin{equation}
\Gamma= \sum_{n}
 \Gamma^{i_1} \cdots \Gamma^{i_n} F_{i_i \cdots i_m} (X^{a'}, \theta^{\alpha'},
h_{ijk} )
\end{equation}
All terms with even number $n=2,4,6 $ of 6 $\Gamma^{i}$ will contribute only to
$C$,
all terms with odd number $n=1,3,5$  of  $\Gamma^{i}$ will contribute to $A$
since
$\Gamma^{i}$ is off-diagonal in our basis. The dependence on diagonal matrices
$\Gamma_{a'}$ is included in $F$. Thus
\begin{equation}
\Gamma = \Gamma_C +\Gamma_A =  \pmatrix{C&
0\cr
\cr               0 &- C} +  \pmatrix{0&
(1- C^2) A^{-1}\cr
\cr                A &0}\,  \, ,
\end{equation}
where
\begin{equation}
\Gamma_C  \equiv  \pmatrix{C&
0\cr
\cr               0 &- C} = \sum_{n=2,4,6}
 \Gamma^{i_1} \cdots \Gamma^{i_n} F_{i_i \cdots i_m} (X^{a'}, \theta^{\alpha'},
h_{ijk}, a )
  \, ,
\end{equation}
and
\begin{equation}
\Gamma_A  \equiv   \pmatrix{0&
(1- C^2) A^{-1}\cr
\cr                A &0}\,  =\sum_{n=1,3,5}
 \Gamma^{i_1} \cdots \Gamma^{i_n} F_{i_i \cdots i_m} (X^{a'}, \theta^{\alpha'},
h_{ijk}, a )
  \, ,
\end{equation}
Finally  the 32-dimensional supersymmetry transformation  on the brane is given
by

\begin{equation}
\delta_{\epsilon , \epsilon'}  \theta ^{\alpha'} = - A ^{\alpha'}{}_\beta
[(1+C)^{-1}] ^\beta{}_ \alpha \epsilon ^\alpha + \epsilon^{\alpha'}
\label{susy}\end{equation}

with $A(X^{a'}, \theta^{\alpha'}, h_{ijk}, a )
 $ and $C(X^{a'}, \theta^{\alpha'}, h_{ijk}, a )
 $ presented for the M-5-brane above.

The supersymmetry transformations of the bosonic fields, 5 scalars and a
tensor, can be obtained using  the combination of $\kappa$-symmetry and
space-time supersymmetry of these fields and the expression (\ref{susy}).
The linearized form of the worldvolume supersymmetry of the (0,2) tensor
multiplet was given in \cite{HSW}.  In notation appropriate to a 6-dimensional
theory for $\epsilon'=0$
\begin{equation}
\delta_{\epsilon}  \theta ^s_{\hat \beta} = \epsilon ^{\hat \alpha t} \Bigl
({1\over 2}
\sigma^i _{\hat \alpha \hat \beta} (\gamma_{b'})_t{}^s \partial_i X^{b'} -
{1\over 6} (\sigma^{ijk}_{\hat \alpha \hat \beta} \delta_t{}^s h_{ijk}\Bigr)
\end{equation}
One can recognize here terms linear and cubic in $\Gamma^i$ which form the
linear approximation of our matrix $A$.

Note however that the full non-linear action of the self-interacting tensor
multiplet has also a symmetry under additional 16-component chiral spinor
$\epsilon^{\alpha'}= \epsilon_{\hat \alpha s}$.  The one with the anti-chiral
spinor
$\epsilon^{\alpha}= \epsilon^{\hat \alpha}_s $   in the linear approximation
relates the spinor of the tensor multiplet to the derivative of  scalars and to
the tensor field strength. The non-linear action has both chiral as well as
anti-chiral supersymmetries.

\section{Conclusion}

We have presented here new possibilities to gauge-fix $\kappa$-symmetry
which may be useful in the context of  {\it new new generation of
$\kappa$-symmetric actions}. In particular  the most recent  $\kappa$-symmetric
theory  describing an $SL(2, {\bf Z})$ covariant Superstring \cite{ct} may need
for quantization a Killing spinor of the background to keep the $SL(2, {\bf
Z})$ symmetry of the quantized theory.
The main emphasis of the new quantization is to take into account the Killing
spinors of the background for making $\kappa$-symmetry irreducible and for
projecting out  1/2 of the space-time fermions. This leads in particular to a
natural  construction of  globally  supersymmetric theories on the worldvolume.
Our main general result is for the space-time Killing spinor adapted gauges,
given the $\kappa$-symmetry generator
$$\Gamma = \pmatrix{
C & (1-C^2)A^{-1} \cr
A  & -C \cr
} \ ,$$
 the global supersymmetry on the world volume is

$$
\delta \theta  = - A (1+C)^{-1} \epsilon  + \epsilon' \ .
$$

As an application of this new quantization we have gauge-fixed the M-5-brane
action in a Killing spinor adapted gauge and obtained the non-linear action and
non-linear supersymmetry of the self-interacting  (0,2) tensor multiplet in
d=6. A further study of this theory
will be necessary  to present a more detailed and explicit structure of it.
Even more effort may be required to construct the interaction of $n$ of such
tensor multiplets.

\vskip 1 cm
I am grateful to E. Bergshoeff for the valuable comments to this work and
information on d=6 tensor multiplets.
  This work   is
supported by the NSF grant PHY-9219345.


\newpage

\begin{thebibliography}{30}



\bibitem{Ce1} M.~Cederwall, A.~von Gussich, B.E.W.~Nilsson,
and A.~Westerberg Nucl. Phys.  {\bf B490},  163 (1997); M.~Cederwall, A.~von
Gussich, B.E.W.~Nilsson, P.~Sundell and A.~Westerberg, Nucl. Phys.  {\bf B490},
 179 (1997).

\bibitem{Ag1}  M.~Aganagic, C.~Popescu, and J.H.~Schwarz,  Phys. Lett. {\bf
B393},  311 (1997), Nucl.~Phys.~{\bf B495},  99 (1997).

\bibitem{Be1} E.~Bergshoeff and P.K.~Townsend,   Nucl.~Phys.~{\bf B490},  145
(1997), hep-th/9611173.

\bibitem{5b} P. Pasti, D. Sorokin and M. Tonin, Phys. Lett.  {\bf 398B}
(1997) 41; I. Bandos, K. Lechner, A. Nurmagambetov, P. Pasti, D. Sorokin and M.
Tonin,  Phys. Rev. Lett. {\bf 78} (1997) 4332.

\bibitem{ag} M. Aganagic, J. Park, C. Popescu and J. H. Schwarz,  Nucl. Phys.
{\bf B496} (1997) 191.
\bibitem{kallosh} R. Kallosh, {\it Covariant quantization of D--branes},
hep--th/9705056.

\bibitem{bk} E. Bergshoeff, R. Kallosh, T. Ort\'\i n and G. Papadopulos, {\it
$\kappa$-symmetry, Supersymmetry and Intersecting Branes}, hep-th/9705040.

\bibitem {Romans1} L.J.~Romans, Nucl. Phys. {\bf B269} (1986) 691.

\bibitem{BSS} E.~Bergshoeff, E. Sezgin and E. Sokatchev, Class.Quant.Grav. {\bf
13}  (1996) 2875, hep-th/9605087.
\bibitem{BFSS} T. Banks, W. Fischler, S. Shenker and L. Susskind,
 Phys. Rev.{\bf D55}, (1997), 112.

\bibitem{S} M.Berkooz, M.Rozali, N.Seiberg, {\it Matrix description of M theory
on $T^4$ and $T^5$},
hep-th/9704089; N.Seiberg, {\it Note on theories with 16 supercharges},
hep-th/9705117


\bibitem{HSW}P.S. Howe and E. Sezgin, Phys. Lett. {\bf B394} (1997) 62,
hep-th/9611008; P.S. Howe, E. Sezgin and P.C. West, Phys. Lett. {\bf B399}
(1997) 49, hep-th/9702008; P.S. Howe, N.D. Lambert and P.C. West, {\it The
self-dual string soliton}, hep-th/9709014.


\bibitem{st} D. Sorokin and  P.K.~Townsend {\it M-theory superalgebra from the
M-5-brane}, hep--th/9708003.

\bibitem{duality} R. Kallosh,  Phys. Rev. D{\bf 52}, 6020-6042 (1995).

\bibitem{equiv} I. Bandos, K. Lechner, A. Nurmagambetov, P. Pasti, D. Sorokin
and M. Tonin, {\it On the equivalence of different formulations of the M theory
five-brane} hep-th/9703127.

\bibitem{ct} M.~Cederwall and P.K.~Townsend, {\it The manifestly $SL(2,{\bf
Z})$-covariant superstring}, \\
hep-th/9709002.

\end{thebibliography}
\end{document}